\documentclass{article}

\usepackage{arxiv}

\usepackage[utf8]{inputenc} 
\usepackage[T1]{fontenc}    
\usepackage{hyperref}       
\usepackage{url}            
\usepackage{booktabs}       
\usepackage{amsfonts}       
\usepackage{nicefrac}       
\usepackage{microtype}      
\usepackage{lipsum}		
\usepackage{graphicx}
\usepackage{natbib}
\usepackage{doi}
\usepackage{booktabs} 
\usepackage{colortbl} 
\usepackage{xcolor}   
\usepackage{caption}  

\definecolor{customgray}{gray}{0.95}

\title{An experiment on an automated literature survey of data-driven speech enhancement methods}

\author{ \href{https://orcid.org/0000-0002-3989-7105}{\includegraphics[scale=0.06]{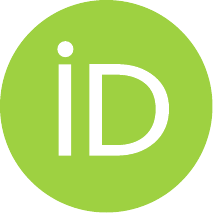}\hspace{1mm}Arthur dos Santos} \\
	Universidade Estadual de Campinas\\
        Campinas, Brazil \\
	\texttt{a264372@dac.unicamp.br} \\
	\And
	\href{https://orcid.org/0000-0001-5478-438X}{\includegraphics[scale=0.06]{orcid.pdf}\hspace{1mm}Jayr Pereira} \\
	NeuralMind\\
	Campinas, Brazil \\
	\texttt{jayr.pereira@neuralmind.ai} \\
        \And
	\href{https://orcid.org/0000-0002-2600-6035}
    {\includegraphics[scale=0.06]{orcid.pdf}\hspace{1mm}Rodrigo Nogueira} \\
	NeuralMind\\
	Campinas, Brazil \\
	\texttt{rodrigo.nogueira@neuralmind.ai} \\
    \And
	\href{https://orcid.org/0000-0002-2246-4450}
    {\includegraphics[scale=0.06]{orcid.pdf}\hspace{1mm}Bruno Masiero} \\
	Universidade Estadual de Campinas\\
        Campinas, Brazil \\
	\texttt{masiero@unicamp.br} \\
    \And
	\href{https://orcid.org/0000-0001-6551-9784}
    {\includegraphics[scale=0.06]{orcid.pdf}\hspace{1mm}Shiva Sander-Tavallaey} \\
	ABB Corporate Research, Västerås, Sweden\\
        KTH Royal Institute of Technology, Stockholm, Sweden \\
	\texttt{shiva.sander-tavallaey@se.abb.com, tssander@kth.se} \\
    \And
	\href{https://orcid.org/0000-0001-5723-9571}
    {\includegraphics[scale=0.06]{orcid.pdf}\hspace{1mm}Elias Zea} \\
        KTH Royal Institute of Technology\\
        Stockholm, Sweden \\
	\texttt{zea@kth.se} \\	
}

\renewcommand{\shorttitle}{An experiment on an automated literature survey of data-driven SE methods}

\hypersetup{
pdftitle={An experiment on an automated literature survey of data-driven speech enhancement methods},
pdfauthor={Arthur dos Santos, Jayr Pereira, Rodrigo Nogueira, Bruno Masiero, Shiva Sander-Tavallaey, Elias Zea},
pdfkeywords={speech enhancement methods, data-driven acoustics, literature survey, natural language processing, large language models},
}

\begin{document}
\maketitle

\begin{abstract}
The increasing number of scientific publications in acoustics, in general, presents difficulties in conducting traditional literature surveys. This work explores the use of a generative pre-trained transformer (GPT) model to automate a literature survey of 116 articles on data-driven speech enhancement methods. The main objective is to evaluate the capabilities and limitations of the model in providing accurate responses to specific queries about the papers selected from a reference human-based survey. While we see great potential to automate literature surveys in acoustics, improvements are needed to address technical questions more clearly and accurately.
\end{abstract}


\section{Introduction}\label{sec:1}

\begin{figure}[ht!]
\centering
\includegraphics[width=0.7\linewidth]{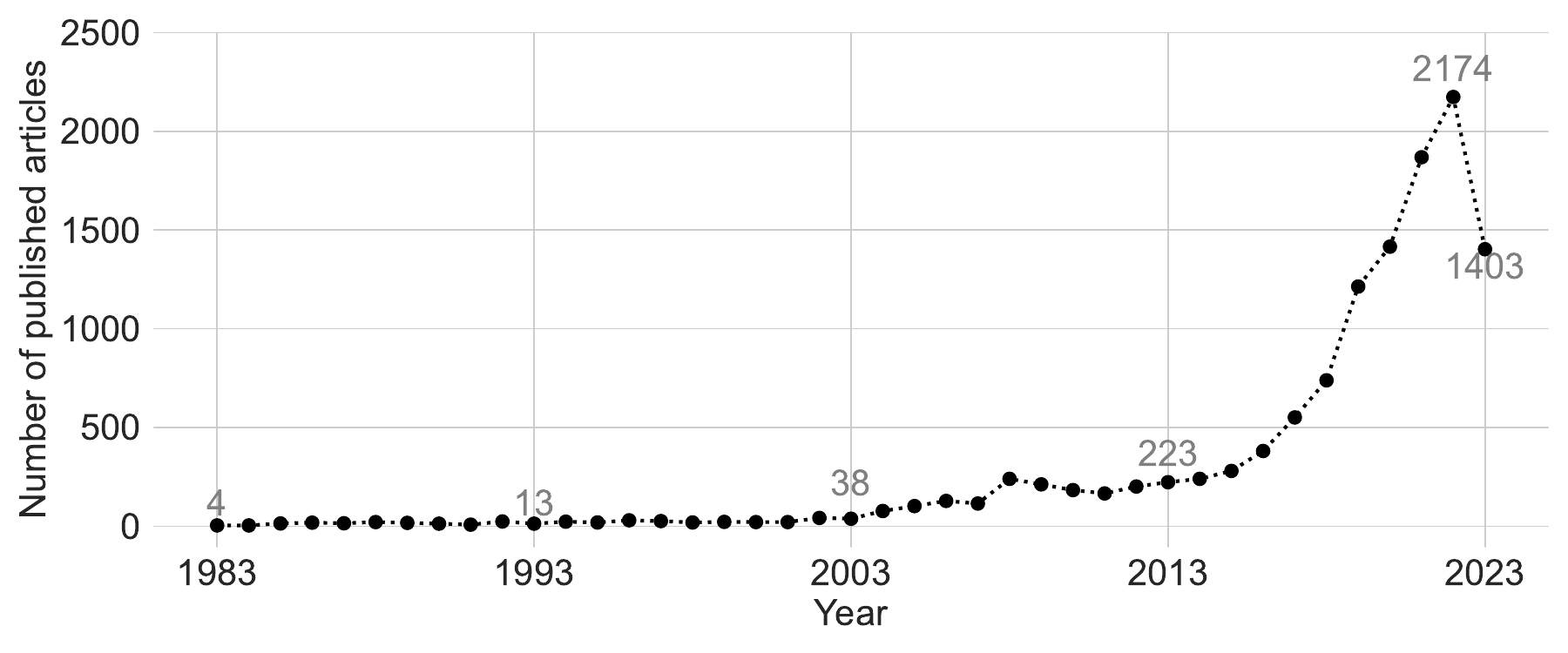}
\caption{Four decades of articles on applications of data-driven methods in acoustics. The results have been obtained from a Scopus search on August 2, 2023.}
\label{fig:publishrates}
\end{figure}

A recent study has shown an increasing publication rate after analyzing 45 million scientific articles produced in the past six decades ~\citep{Park2023}. In the context of applications of data-driven methods in acoustics alone, as shown in the Scopus\footnote{\url{https://www.scopus.com/}} search in Fig.~\ref{fig:publishrates}, the number of articles in the first half of 2023 had exceeded the total number of articles in the entire year of 2019. Given this growth in the literature, the acoustics community faces the limitations of traditional survey methods. 
At the same time, the remarkable advancements in the field of natural language processing (NLP) and large language models (LLMs) in recent years---leading to the ``boom'' of the generative pre-trained transformer (GPT)~\citep{Stokel-Walker2023}, offers a unique opportunity to guide and advance knowledge in acoustics through automated large-scale text processing. This can provide more accessible information for researchers, practitioners, and engineers interested in data-driven methods for acoustics and vibration in the broader sense.

Recent literature surveys in acoustics 
have reviewed the theory and applications of machine learning (ML) in acoustics~\citep{Bianco2019}, sound source localization (SSL) using deep learning methods~\citep{Grumiaux2022}, as well as noise-induced hearing loss in several contexts~\citep{Neitzel2019, Radziwon2019, Malowski2022}. The survey by Gannot et al. analyzed $393$ papers on speech enhancement and source separation through four queries~\citep{Gannot2017}: what is the acoustic impulse response model, what is the spatial filter design, what is the parameter estimation algorithm, and what is the post-filtering technique? Other related review papers have covered more specific applications of acoustics, such as source-range estimation for underwater acoustics~\citep{Song2022}, SSL for wireless acoustic sensor networks~\citep{Cobos2017}, the LOCATA challenge for source localization and tracking~\citep{Evers2020}, and 15 years of SSL in robotics applications~\citep{Argentieri2015}.

Writing a literature survey can be viewed as the art of \emph{making a long story short}, which can be pretty laborious. Typically, it starts by selecting a topic of interest and elaborating a list of questions. Then, a search for relevant literature items must be fulfilled, which, nowadays, can be facilitated by search engines and databases that assess the credibility and reliability of sources (e.g., Scopus, Google Scholar,\footnote{\url{https://scholar.google.com/}} etc.). 
This is followed by processing the selected literature, organizing items into categories based on their similarities and differences, analyzing them, and noting essential trends, patterns, knowledge gaps, etc. To do so, several tools exist to provide researchers with ways to document the whole process, with mechanisms to build quality assessment checklists, data extraction forms, among others (e.g., Covidence,\footnote{\url{https://www.covidence.org/}} Parsif.al,\footnote{\url{https://parsif.al/}} Rayyan,\footnote{\url{https://www.rayyan.ai/}} etc.). However, until now, one has to \textit{read through} all the literature.

Reading a scientific paper typically involves scanning the text for the research problem, assumptions, methods, evaluations, and main findings; interpreting relevant mathematical terminology; understanding the structure and organization of the text; and synthesizing information to form a coherent understanding of it as a whole~\citep{Pain2016}. Thus, the time taken to read an academic paper varies depending on various factors, such as its length, the complexity of the topic, and the reader's familiarity with the subject matter. Assuming that a familiar reader has a typical reading speed of approximately $200$–$300$ words per minute~\citep{Frank1990}, it would take roughly $1$–$2$ hours to read a $10$-page academic paper. Math-intensive documents might take even longer. Therefore, scanning 100 articles would take approximately one month of uninterrupted work to read through the literature.

The usage of LLMs for automated text summarization and generation is relatively new and has had applications in medicine and news enterprises. A relevant study to this work was published recently by Tang et al.~\citep{Tang2023}, who performed zero-shot medical evidence summarization generated with GPT-3.5 and ChatGPT and compared them to human-generated summarization. Similar methodologies have been applied to, for example, compare abstracts generated by ChatGPT to real abstracts from medical journals~\citep{Gao2023}, identify and assess key research questions in gastroenterology~\citep{Lahat2023}, and answer multiple-choice questions about human genetics~\citep{Duong2023}. LLMs have also been used for automatic news summarization~\citep{Syed2020,Goyal2023}. A common element in these studies is that LLM-based methodologies have substantial potential in medical and news applications, but more work is needed to increase the accuracy and fidelity. 

In this paper, we employ a GPT model to query a literature corpus comprising $116$ texts on data-driven speech enhancement methods. The main goal is to speed up literature surveys. The structure of this paper is as follows: Section~\ref{sec:method} describes the methodology, including the literature corpus, a short description of the GPT model, and the queries posed to the model. Section~\ref{sec:results} presents the results of the GPT model and a comparison with a reference (human-based) survey~\citep{dosSantos2022}. Lastly, conclusions are drawn in Sec.~\ref{sec:conclude}.

\section{Methodology}
\label{sec:method}

\subsection{Text corpus}

In this study, the corpus consists of 116 articles published in the English language between January and December $2021$, matching the search strings “\textit{audio enhancement}” OR “\textit{dereverberation}” AND in the context of “\textit{machine learning}” OR “\textit{deep learning},” from various databases, including the AES E-Library,\footnote{\url{https://www.aes.org/e-lib/}} ACM Digital Library,\footnote{\url{https://dl.acm.org/}} Google Scholar, IEEE Digital Library,\footnote{\url{https://ieeexplore.ieee.org/}} JASA,\footnote{\url{https://asa.scitation.org/journal/jas}} MDPI,\footnote{\url{https://www.mdpi.com/}} ResearchGate,\footnote{\url{https://www.researchgate.net/}} Research Square,\footnote{\url{https://www.researchsquare.com/}} ScienceDirect,\footnote{\url{https://www.sciencedirect.com/}} Springer,\footnote{\url{https://link.springer.com/}} arXiv,\footnote{\url{https://arxiv.org/}} and some repositories of higher education institutions and subsidiary research departments of corporations.

Conference, journal, and challenge papers, book series and chapters, extended abstracts, technical notes, M.Sc. theses, and Ph.D. dissertations were included in the search. The average number of pages per article was $8$, varying from $2$ to $30$ (except for the M.Sc. and Ph.D. monographies, which varied from $31$ to $118$). For the complete list of texts reviewed, please refer to this external 
{link}\footnote{\url{https://drive.google.com/file/d/1rpRiSjyNpHIF9GzNzy8qTQEmKHLMzDkN/}}.

\subsection{Generative pre-trained transformer model}

First released in 2018~\citep{Radford2018} and then continuously updated, the generative pre-trained transformer (GPT) is a large autoregressive language model designed to generate human-like responses to natural language input. It can be used for various tasks, including chatbots, language translation, and text summarization. Its ability to generate coherent text has made it a valuable tool for researchers and developers in NLP and ML applications. For example, it is possible today to ask ChatGPT or Bard to summarize a scientific paper or generate a list of sources for a literature survey on a specific topic. However, it has been seen that generated responses are often partially (sometimes entirely) fake~\citep{Alkaissi2023}, and the answering accuracy can deteriorate when the answer to the query lies in the middle of the context~\citep{Liu2023}. Therefore, we have focused on applying the underlying GPT model, not on the direct usage of chatbots. 

In this study, we used the large language model of Open AI \texttt{GPT3.5-turbo-16k}\footnote{\url{https://platform.openai.com/docs/models/gpt-3-5}} to process the research papers and extract relevant information. This allows us to explore the model's ability to handle long contexts (i.e., 16k tokens or up to about 50 pages of pure text, assuming an average of 300 tokens/page), enabling a comprehensive analysis of an entire scientific paper. This is in contrast to previous studies on automatic literature summarization~\citep{Tang2023}, which examined scientific abstracts. It should be stressed that articles in PDF in acoustics most often translate into fewer pages of pure text due to figures, tables, etc. We utilize the GPT model's ability to answer questions to help us address specific inquiries about the papers. 
First, we convert the PDF files into text using the PyPDF2 library\footnote{\url{https://pypi.org/project/PyPDF2/}}. Next, we prompt the GPT model with each paper's full text and specific questions to obtain comprehensive answers. This iterative process is performed for every paper to address the four queries presented in the following section. Compared to the human pace, this methodology requires much less time to analyze academic papers and provide an answer to the question posed. 

\subsection{Queries}
\label{sec:queries}

Four questions are considered in this study: two relatively ``simple'' and two relatively ``hard'':
\begin{itemize}
    \item \textbf{Query 1 (Q1)}: What country were the authors based in? The output of this question is a list of the authors' countries of affiliation. 
    \item \textbf{Query 2 (Q2)}: Was it single-channel or multi-channel scenario? The output of this question is either one of the two classes (single or multi), and we are interested in determining the probability of the GPT obtaining the class right.  
    \item \textbf{Query 3 (Q3)}: What type of architecture was used? This question is relatively more difficult than the previous one, requiring domain knowledge for proper comprehension. This question relates to determining the data-driven model used in the studies. Thus, the output of this question is a string, and we are interested in knowing the probability that the GPT will produce the string as accurately as possible. 
    \item \textbf{Query 4 (Q4)}: In what context were these applications used (e.g., hearing aids, communication, speech enhancement)? This is the most challenging question posed to the GPT in this study, which involves determining the application area of speech enhancement considered in previous studies. Thus, the output of this question is a string, and we are interested in determining the probability that the GPT will produce the string as accurately as possible.  
\end{itemize}

\section{Results}
\label{sec:results}

\subsection{Outputs from questions}

These four questions were selected from our reference literature survey~\citep{dosSantos2022}, whose answers are taken as ground truth. Section~\ref{sec:humanbased} summarizes the answers presented in \citep{dosSantos2022}, whereas Sec.~\ref{sec:gptbased} compares the answers produced by the GPT model with the answers in the reference survey. 

\subsubsection{Human-based survey}
\label{sec:humanbased}

\begin{figure}[h!]
\centering
\includegraphics[width=0.75\linewidth]{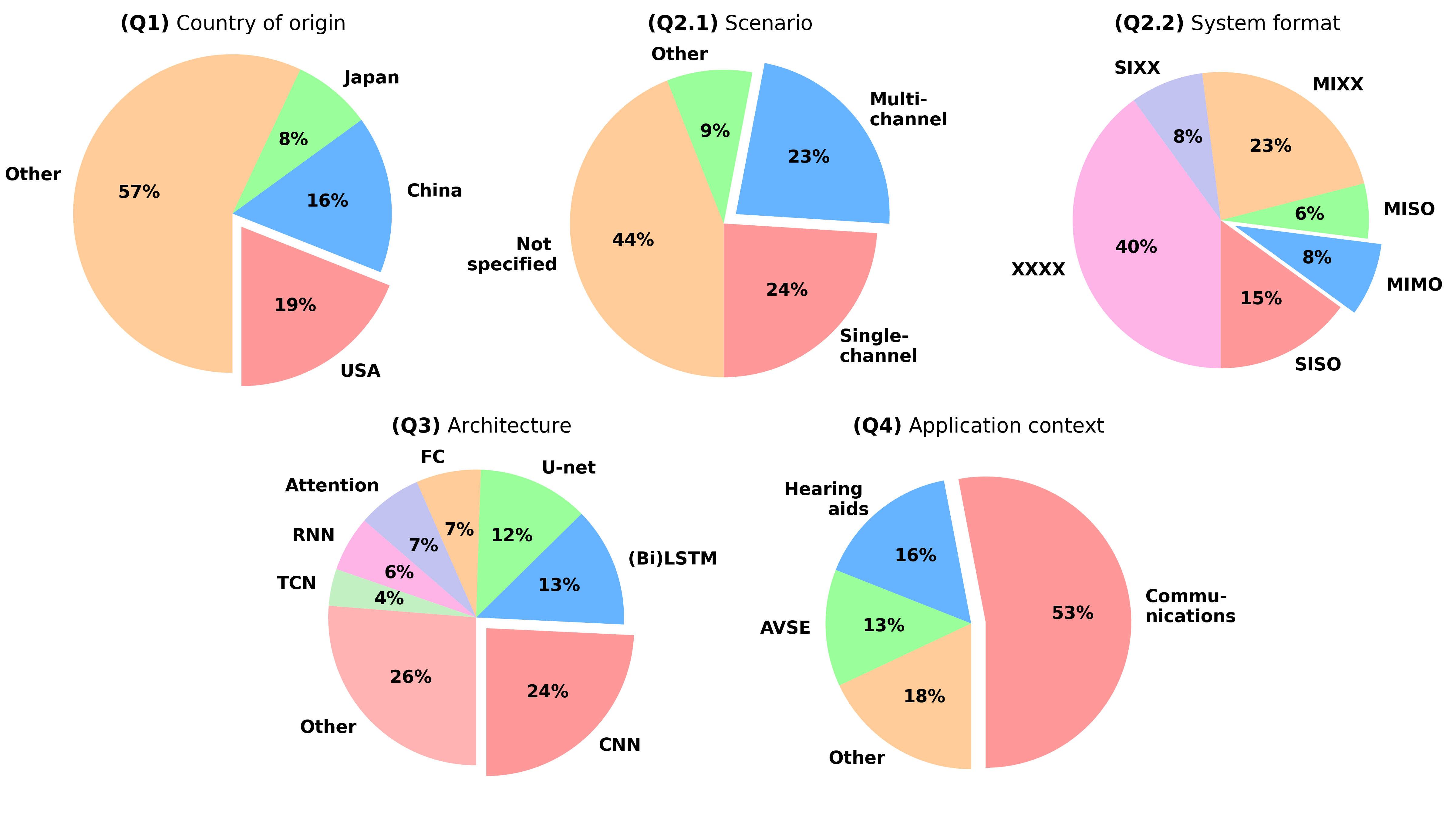}
\caption{Simplified pie-charts for the human-based survey~\citep{dosSantos2022}.}
\label{fig:piecharts}
\end{figure}

Authors' affiliations include higher education institutions, subsidiary research departments of corporations (e.g., Adobe, Facebook, Google, Microsoft), and semi-private and fully financed government research institutions. The main contributors were the United States of America (USA), China, and Japan, as illustrated in Figure~\ref{fig:piecharts} (Q1), with 28 countries represented. Other contributing countries include South Korea, Germany, the United Kingdom (UK), India, Switzerland, France, Denmark, the Netherlands, Canada, Ireland, Italy, Norway, Spain, Taiwan, Vietnam, Austria, Brazil, Chile, Greece, Hong Kong, Israel, Malaysia, Pakistan, Poland, and Singapore.

Not all corpora account for multi-channel scenarios. Among the reviewed articles, only $23\%$ explicitly addressed multi-channel scenarios, whereas $24\%$ focused on single-channel scenarios, as illustrated in Figure~\ref{fig:piecharts} (Q2.1). Other scenarios include binaural, Ambisonics, and stereo signals. However, most articles did not specify this information. For articles with a complete system format or configuration details, most are single-input-single-output (SISO) systems, followed by multiple-input-multiple-output (MIMO) and multiple-input-single-output (MISO) systems, as shown in Figure~\ref{fig:piecharts} (Q2.2). Other formats include multiple-input systems without a specified output format (MIXX), single-input systems without a specified output format (SIXX), and systems with completely unspecified input-output formats (XXXX).

The most commonly used model architectures are $1$-D and $2$-D Convolutional Neural Networks (CNN), uni- or bi-directional Long Short-Term Memory (LSTM) blocks, U-net, Fully Connected (FC) architectures, attention networks, recurrent neural networks (RNN), and temporal convolutional networks (TCN), as illustrated in Figure~\ref{fig:piecharts} (Q3). Other architectures include adversarial, convolutional, encoder/decoder, feedforward, geometrical, neural beamformer, recurrent, reinforcement learning, Seq2Seq, and statistical/probabilistic models.

Applications are often joint, including speech enhancement, dereverberation, noise suppression, speech recognition, and source separation. These applications focus mainly on communication, hearing aids, and audio-visual speech enhancement (AVSE), as illustrated in Figure~\ref{fig:piecharts} (Q4). Additional applications include suppressing nonlinear distortions, enhancing heavily compressed signals in speech and musical domains, audio inpainting applied to both speech and music signals, law enforcement and forensic scenarios, acoustic-to-articulatory inversion, input-to-output mapping of auditory models, studio recordings, and selective noise suppression.

\subsubsection{Machine-based survey}
\label{sec:gptbased}

Because the GPT model is designed to generate human-like responses to natural language input, even if prompted with the same questions posed by humans, its answers are expected to vary from those of humans. To quantify the extent to which these variations differ from the desired responses, a tier list was elaborated as follows to compare the machine-based results with the human ground truth:

\begin{itemize}
    \item \textbf{Tier 1} - No answer / Completely wrong / Not a pertinent answer: the model fails to provide any response or provides a completely incorrect or irrelevant answer (e.g., the author failed to mention, yet GPT prompts a specific answer);
    \item \textbf{Tier 2} - Marginally correct: the model provides a response that contains at least some correct information;
    \item \textbf{Tier 3} - Mostly correct with minor errors or omissions: the model produces the majority of the information correctly but might miss a few details or make minor mistakes;
    \item \textbf{Tier 4} - Perfectly correct: the model produces completely accurate and correct responses.
\end{itemize}

The first author, who also conducted the reference human-based survey, performed the tier-based assessment of responses to the survey questions (Q1–Q4 in Sec. ~\ref{sec:queries}). This choice has been taken to prevent the need for an analysis of the subjective interpretation of the machine-generated responses, which is beyond the scope of this paper. It is worth noting that evaluating more technical questions, such as Q3 and Q4, requires domain knowledge of speech enhancement and data-driven methods. However, assessing responses to more straightforward questions like Q1 and Q2 requires little to no domain knowledge. 

\begin{figure}[h!]
\centering
\includegraphics[width=0.75\linewidth]{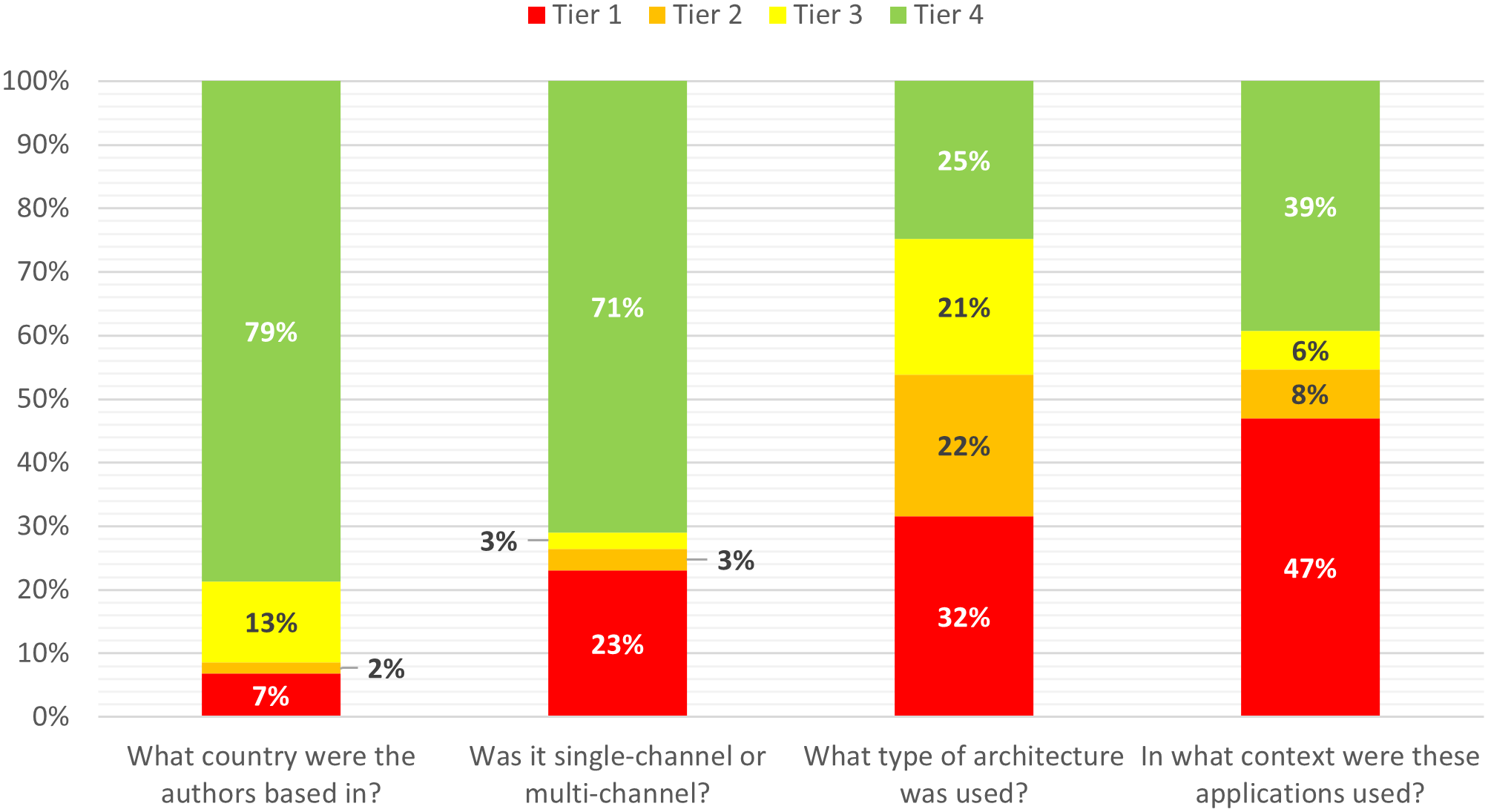}
\caption{Stacked bar charts for the machine-based answers to the four questions using the corpus of 116 papers.}
\label{fig:barchart}
\end{figure}

Figure~\ref{fig:barchart} illustrates the stacked bar charts containing the tier distribution for each question after comparing the machine-based responses with the human-based responses. For full results of the raw human-based survey in comparison with machine outputs for Q1-Q4, please refer to this external {link}\footnote{\url{https://drive.google.com/drive/folders/1jfud4LVkwBQd8KhWKHxXUAkaxZCO9igu?usp=sharing}}. In what follows, the results are analyzed in more detail. 

\subsection{Analysis of results}

From Fig.~\ref{fig:barchart}, most answers are either perfectly correct or have minor errors for Q1 (``\textit{What country were the authors based in?}''), which is a simple question that can be answered based on authors’ affiliations. In this case, the errors could be related to the fact that the country of affiliation was not included or correctly linked to their names in the provided metadata.

Table~\ref{tab:q1} illustrates examples of human-based ground-truth versus machine-based responses for Q1. It can be seen that some machine-based answers are more concise than others, specifically, stating only the country versus starting a sentence with ``The authors were based in...'' This reflects the coherent and diverse capacity of the GPT model to respond to such a question. We strongly suspect that the model's accuracy can be improved by providing only the article's metadata as context instead of the complete text, thus minimizing potential issues due to the length of the context~\citep{Liu2023}. 

\begin{table}[!ht]
\centering
\scriptsize
\caption{Examples of human-based ground truth versus machine-based responses for Q1.}
\label{tab:q1}
\vskip3pt
\begin{tabular}{>{\centering\arraybackslash}p{1cm} | >{\centering\arraybackslash}p{4.5cm} | >{\centering\arraybackslash}p{6.5cm}}
\toprule
\textbf{} & \textbf{Human-based ground truth} & \textbf{Machine-based responses} \\
\midrule
\rowcolor[gray]{.95}\textbf{Tier 1} & Germany, UK & The authors were based in the USA and China. \\
\textbf{Tier 2} & China, USA, Denmark & China \\
\rowcolor[gray]{.95}\textbf{Tier 3} & Germany, Canada & Germany \\
\textbf{Tier 4} & UK, Switzerland & The authors were based in the UK (specifically, the University of Edinburgh) and Switzerland (specifically, ETH Zurich). \\
\bottomrule
\end{tabular}
\end{table}

Regarding Q2 (``\textit{Was it single-channel or multi-channel scenario?}''), most predictions are perfectly correct; however, there is an increase in completely inaccurate answers compared to Q1. This is partially due to cases with no specified response (the authors failed to mention) and the GPT model prompting a response~\citep{Alkaissi2023} instead of stating that it could not find that information in the text provided. In addition, there are cases where the speech enhancement method could be used for multi-channel purposes, yet the authors chose not to. Interestingly, the GPT model may assign these cases as single- and multi-channel scenarios. 

Table~\ref{tab:q2} illustrates examples of human-based ground-truth versus machine-based responses for Q2. The examples show that the GPT model provides nuanced answers for the different tiers, including ``The proposed system includes...'' or ``The approach described in the paper is...'' This reflects the complexity of the text generation capabilities of the GPT used in this experiment. As pointed out in Sec.~\ref{sec:queries}, the prompt is a classification question: whether the approach is of a single-channel or multi-channel class. However, when looking at the Tier 2 example in Table~\ref{tab:q2}, the authors discussed a single-channel approach [see Ref. 107 in~\citep{dosSantos2022}]; however, it can also be applied in multi-channel scenarios, which is the machine response. Similarly, for Tier 3, the GPT model does not explicitly produce the result ``Binaural'' or ``Monoaural''. However, it responds that the approach belongs to the multi-channel class. Making the question more precise may help to improve the model's accuracy further.

\begin{table}[!ht]
\centering
\scriptsize
\caption{Examples of human-based ground truth versus machine-based responses for Q2.}
\label{tab:q2}
\vskip3pt
\begin{tabular}{>{\centering\arraybackslash}p{1cm} | >{\centering\arraybackslash}p{4.5cm} | >{\centering\arraybackslash}p{6.5cm}}
\toprule
\textbf{} & \textbf{Human-based ground truth} & \textbf{Machine-based responses} \\
\midrule
\rowcolor[gray]{.95}\textbf{Tier 1} & Not specified & The approach described in the paper was a single-channel approach. \\
\textbf{Tier 2} & Single-channel & The proposed system includes both single-channel and multi-channel approaches. \\
\rowcolor[gray]{.95}\textbf{Tier 3} & Binaural, Monaural & The proposed method is a multichannel approach. \\
\textbf{Tier 4} & Multi-channel & The approach described in the paper is a multi-channel approach. \\
\bottomrule
\end{tabular}
\end{table}

For question Q3 (``\textit{What type of architecture was used?}''), there is an observable balance between all tiers. One of the most common reasons for completely wrong answers is that the GPT model identifies the name of the ``trade'' architecture as the type of architecture (e.g., ``VGGNet'' instead of ``fully connected, CNN''). We suspect this can be improved by fine-tuning the GPT model to determine the underlying architecture instead of its variant name. Another common error is simply outputting the answer ``DNN'' (e.g., ``the network architecture is a DNN'') instead of detailing its type. Once again, we strongly suspect that providing the GPT model with the necessary context would prevent these mistakes. At any rate, most answers are partially correct, i.e., either it got something or almost everything right, which, together with the wrong answers, reduces the quantity of perfectly correct answers. 

Table~\ref{tab:q3} presents examples of human-based ground-truth predictions versus machine-based predictions for Q3. Interestingly, for Tier 3, it can be seen that the GPT model not only (nearly) produces the right architecture type of CNN, but it also adds ``variable dilation factors.'' Based on our observations with other papers on the survey, this extracted additional information, if accurate, holds significant value and analysis depth in the context of large-scale surveys. 

\begin{table}[!ht]
\centering
\scriptsize
\caption{Examples of human-based ground truth versus machine-based responses for Q3.}
\label{tab:q3}
\begin{tabular}{>{\centering\arraybackslash}p{1cm} | >{\centering\arraybackslash}p{4.5cm} | >{\centering\arraybackslash}p{6.5cm}}
\toprule
\textbf{} & \textbf{Human-based ground truth} & \textbf{Machine-based responses} \\
\midrule
\rowcolor[gray]{.95}\textbf{Tier 1} & Fully connected, LSTM & The network architecture used in the proposed method is a deep neural network (DNN). \\
\textbf{Tier 2} & Attention Network, Autoencoder, Conformer Network, CNN, LSTM & The network architecture used in the study is a conformer-based architecture. \\
\rowcolor[gray]{.95}\textbf{Tier 3} & CNN, RNN & The network architecture used in the study is a convolutional neural network (CNN) with variable dilation factors. \\
\textbf{Tier 4} & CNN & The network architecture used in the study was Convolutional Neural Networks (CNNs). \\
\bottomrule
\end{tabular}
\end{table}

Finally, for question Q4 (``\textit{In what context were these applications used?}''), most answers were perfectly correct or entirely incoherent. This is because, in most cases, the authors do not mention the context of their applications in the full texts. This adds a higher degree of complexity to the GPT model to infer the application from partially incomplete information, something a human with domain knowledge might infer more accurately at this point. Still, it is interesting that the GPT model considers the broader field of study (e.g., dereverberation and speech enhancement) as an application context and attempts to answer the query nonetheless. Table~\ref{tab:q4} illustrates examples of human-based ground-truth versus machine-based predictions for Q4. As can be seen, the Tier 4 response example is remarkably similar to the human response. However, for Tier 1, the GPT model answers even though the human has found that it is not specified in the text. Further examination and understanding must rely on the GPT model to answer these queries more accurately. 

\begin{table}[ht!]
\centering
\scriptsize
\caption{Examples of human-based ground truth versus machine-based responses for Q4.}
\label{tab:q4}
\begin{tabular}{>{\centering\arraybackslash}p{1cm} | >{\centering\arraybackslash}p{4.5cm} | >{\centering\arraybackslash}p{6.5cm}}
\toprule
\textbf{} & \textbf{Human-based ground truth} & \textbf{Machine-based responses} \\
\midrule
\rowcolor[gray]{.95}\textbf{Tier 1} & Not specified & The applications of the Neural Beamforming Front-End described in the paper include real-time audio signal enhancement for human-computer interaction, telecommunication, web conferencing, and pro-audio applications. \\
\textbf{Tier 2} & Voice-Over-IP applications & The applications of high fidelity speech regeneration with application to speech enhancement include speech enhancement for better communication and collaboration, particularly in the context of physical distancing during the COVID-19 pandemic. \\
\rowcolor[gray]{.95}\textbf{Tier 3} & Podcasts, video voice-overs, and audiobooks that require studio-quality audio with full bandwidth and balanced equalization (EQ) & The applications of signal processing to audio and acoustics discussed in the paper are focused on speech enhancement, specifically improving the quality of real-world consumer-grade recordings to sound like studio recordings. This can be beneficial for various content creation tasks such as podcasts, video voice-overs, and audiobooks. \\
\textbf{Tier 4} & Law-enforcement related audio recordings where it's difficult to understand the speech of one specific speaker amongst all the others & The applications mentioned in the text are related to law enforcement and investigations involving audio recordings. \\
\bottomrule
\end{tabular}
\end{table}

\section{Conclusions}
\label{sec:conclude}

In this experiment, we demonstrate the capabilities and limitations of a generative pre-trained transformer (GPT) model to survey a collection of scientific papers on data-driven speech enhancement methods. To the best of the authors' knowledge, this is the first study that examines the use of large language models to automate a literature survey in acoustics. In essence, the GPT model poses four queries to a corpus of $116$ articles, and the machine-generated answers are compared to a human-based ground truth survey. Our findings indicate that simple questions can be answered with significant accuracy. In contrast, more nuanced technical questions require improving the accuracy and clarity of the questions or careful contextualization and fine-tuning of the model. In the future, we hope this paper stimulates the adoption of artificially intelligent systems to aid humans in surveying larger corpora (e.g., thousands of articles) in acoustics. 

\section{Acknowledgments}
This study was partially sponsored by the São Paulo Research Foundation (FAPESP) under grants \#2017/08120-6, \#2019/22795-1, and \#2022/16168-7. We also thank Prof. Roberto Lotufo and Prof. Renato Lopes for their valuable discussions and suggestions. 


\bibliographystyle{unsrtnat}
\bibliography{references}  






\end{document}